# Artificial DNA Lattice Fabrication by Non-Complementarity and Geometrical Incompatibility

*Jihoon Shin,[1] Junghoon Kim,[2] Rashid Amin,[1] Seungjae Kim,[2] Young Hun Kwon,[3]\* and Sung Ha Park,[1,2]\**

[1]Sungkyunkwan Advanced Institute of Nanotechnology (SAINT), Sungkyunkwan University, Suwon 440-746, Korea.

[2]Department of Physics, Sungkyunkwan University, Suwon 440-746, Korea.

[3]Department of Applied Physics, Hanyang University, Ansan 426-791, Korea

\* Address correspondence to sunghapark@skku.edu (SHP) or yyhkwon@hanyang.ac.kr (YHK).

ABSTRACT

Fabrication of DNA nanostructures primarily follows two fundamental rules. First, DNA oligonucleotides mutually combine by Watson-Crick base pairing rules between complementary base sequences. Second, the geometrical compatibility of the DNA oligonucleotide must match for lattices to form. Here we present a fabrication scheme of DNA nanostructures with non-complementary and/or geometrically incompatible DNA oligonucleotides, which contradicts conventional DNA structure creation rules. Quantitative analyses of DNA lattice sizes were carried out to verify the unfavorable binding occurrences which correspond to errors in algorithmic self-assembly. Further studies of these types of bindings may shed more light on the exact mechanisms at work in the self-assembly of DNA nanostructures.

Watson-Crick base pairing, a fundamental concept embodied in DNA nanotechnology,[1,2] is a unique feature that can be used to design and spontaneously self-assemble single strands into tiles and lattice structures in all dimensions.[3-9] Ever since the first realization of DNA nanostructures, there has been a copious amount of research in this field, all which have exploited this particular aspect of complementarity. Here we present a DNA nanostructure fabrication scheme which seems to contradict conventional structure fabrication principles. DNA lattice structures were produced with DNA oligonucleotides having not only non-complimentary sticky ends but also sticky ends which do not conform to helicity or are not antiparallel, *i.e.*, geometrically incompatible designs. In principle, the non-complementarity of this design scheme corresponds to error bindings in DNA algorithmic self-assembly.[10-13] A number of analysis patterns to explain this phenomenon have been devised.

The basic motif used in the fabrication of these lattices was the double-crossover (DX) tile, which consist of two duplexes connected by two Holliday junctions with four 5-nucleotide sticky ends.[14] Non-complementary and geometrically incompatible conditions were applied in contradiction to Watson-Crick base pairing. Two types of samples were prepared. Single tile lattices (STLs), in which the lattice is composed of only one tile type, and double tile lattices (DTLs), in which the lattice is composed of two different types of tiles. Figure 1A, C, E, and G illustrate the geometrically compatible cases while Figure 1B, D, F, and H show geometrically incompatible oligonucleotides which were designed by changing the directionality and helicity of the sticky ends protruding from the DX tiles. Analyses of the STLs were done by obtaining the concentration dependence of the lattice size and the concentration dependence of the coverage, *i.e.*, the fraction of the mica surface covered by the lattice. For the DTLs, we enforced error bindings between the tiles which are equivalent to errors occurring in the one input and one output DNA tile based algorithmic self-assembly.[15] In order to check the characteristics of error bindings, DX tiles with hairpins for different types of DTLs were introduced.

A total of eight different types of DNA samples (four types of STLs and four types of DTLs) were prepared with non-complementary and/or geometrically incompatible sets of oligonucleotides as shown in Figure 1. Four different types of conditions, were applied to both STL and DTL tile types. For the STLs, tiles with both complementarity and geometrical compatibility, denoted as STL(O,O), (Figure 1A), tiles with complementarity and geometrical incompatibility, STL(O,X), (Figure 1B), tiles with non-complementarity and geometrical compatibility, STL(X,O), (Figure 1C), and tiles with non-complementarity and geometrical incompatibility, STL(X,X), (Figure 1D) were prepared. The DTLs comprising the DX tiles can be divided into two different categories depending on the existence of attached hairpin structures which is analogous to the 0/1 bit information tile representation in algorithmic self-assembly. These two tile types are designed in a way that aids the analysis of the binding mechanism between the two DX tiles. Of the two bindings that occur among the three DTL tiles shown in Figure 1E-H, we designed the sticky ends of one of the duplexes so that they always obey complementarity and geometrical compatibility (yellow and red triangular sticky ends in Figure 1E-H). We call such sticky end bindings "common sets" as they are present in all tile types. This ensures that the sticky ends on both ends of one of the duplexes participate in a binding in which complementarity and geometrical compatibility are satisfied. In this regard, unlike the cases of STL(X,O), STL(O,X), and STL(X,X), where all the intentionally designed non-complementary and/or geometrical incompatible bindings are fully satisfied, for DTL(X,O), DTL(O,X), and DTL(X,X), only half of the bindings are non-complementary and/or geometrically incompatible.[16]

## RESULTS AND DISCUSSION

STL images were taken with atomic force microscopy (AFM) to confirm our design under various conditions (Figure 2A-H). For conditions in which lattice formations occurred, two different lattice sizes were found (referred to as sizes **I** and **II**). When complementarity and geometrical compatibility were obeyed (*i.e.*, STL(O,O) and DTL(O,O)), formation of $\sim 1 \times 10^5$ nm$^2$ size (size **I**) lattices occurred which consisted of $2 \times 10^3$ DX tiles. With the exception of STL(X,O), in which no lattice formations occurred, the lattices grown under conditions of either non-complementarity or geometrical incompatibility (*i.e.*, STL(O,X), DTL(O,X), and DTL(X,O)) were found to have sizes of $\sim 2 \times 10^4$ nm$^2$ (size **II**).

To check the lattice size dependence on the DNA tile concentration, samples of 100 nM, 200 nM, 400 nM, 600 nM, and 800 nM were used for the STLs. Although predicting the saturated nucleation concentration (maximum free tile concentration in a test tube) may be difficult from Figure 2I, the saturated lattice concentration (minimum tile concentration for maximum lattice size formation) can be easily determined. For STL(O,O), the saturated lattice concentration was found to be slightly above 400 nM (Figure 2A) with lattices of size **I**. Small patches of remnant lattices persist throughout experiments conducted at increasing DNA concentrations of the annealing process (dashed line in Figure 2I). Although the frequency of lattice formations of size **I** increases slightly with increasing DNA concentrations, the lattice sizes remain almost constant. Figure 2E to H show lattice formations at the highest DNA concentration, 800 nM. AFM images showing the details of the concentration dependence on the lattice size can be found in the supporting information. For STL(O,X), size **II** lattices were observed at DNA concentrations of 100 nM (Figure 8S). The saturated lattice concentration of STL(O,X) is much lower than that of STL(O,O). Figure 2B and F clearly show that the lattice sizes for concentrations of 400 nM and 800 nM are roughly within 10% of each other. In the cases of STL(X,O) and STL(X,X), evidence of lattice structure formation could not be found for any of the DNA concentrations (Figure 2C, D, G and H).

In order to further probe the characteristics of the non-complementary and geometrically incompatible bindings, the concentration dependence of the mica coverage was measured. The coverage of STL(O,O) slightly increases with increasing DNA concentrations (black line in Figure 2J). On the other hand, the coverage of STL(O,X) dramatically increases when the DNA concentration is increased from 200 nM to 400 nM and begins to saturate and completely cover the mica past 400 nM (blue line in Figure 2J). One explanation of this pronounced jump may be due to the low binding energy of the STL(O,X). Since STL(O,X) is geometrically incompatible, the bindings between the sticky ends can be thought to be considerably weaker than that of the bindings which occur for STL(O,O), where complementarity and geometrical compatibility are fully satisfied. This affords for more frequent assembly of lattices smaller than STL(O,O) which would increase the coverage.

To investigate whether more conventional methods of fabricating DNA from multiple DX tiles could be achieved, we applied the same scheme to double tile systems. These systems are similar to the one input and one output logic of DNA tile based algorithmic self-assembled lattices but completely differ in their purpose. Whereas one of the main objectives in lattice growth by algorithmic rules is in the reduction of errors,[17-19] here, we propose the construction of lattices with 100% errors. In all the double tile systems where there is one type of constraint (*i.e.*, DTL(O,X) and DTL(X,O)), the sizes of the lattices were found to be of size **II**. From the AFM data, most of the lattices had ~20 individual layers in the direction of the duplex axis, meaning errors had occurred 20 consecutive times during the growth of the lattice.

In the case of DTL(O,O), one would expect the same formation size as STL(O,O) (Figure 3A) at half the concentration, since under equivalent conditions, DTLs have double the number of DNA strands compared to STLs. Analysis of AFM images revealed this to be true as the lattice sizes for DTL(O,O) and STL(O,O) were the same at concentrations of 200 nM and 400 nM, respectively. As previously mentioned, DTL(O,X), DTL(X,O), and DTL(X,X) consist of two tile types, one with and one without hairpin structures. This double tile design gives rise to three different types of lattice patterns (Figure 3E-G). Yellow rectangles with/without an inner circle represent tiles with/without hairpin structures (1 bit/0 bit). Blue rectangles represent tiles without hairpin structures (0 bit). The green bindings between the two tiles symbolize conditions where complementarity and geometrical compatibility are satisfied and the red bindings symbolize either one or both unsatisfied conditions. Patterns A and B (Figure 3E, F) depict two possible cases in which a lattice is formed from alternating red and green types of bindings. On the other hand, pattern C (Figure 3G) is a random mixture of patterns A and B. Interestingly, only pattern B lattices were found for DTL(O,X), whereas all three lattice patterns were found for DTL(X,O) with an occurrence rate of 10%, 70%, and 20% for patterns A, B, and C, respectively (Figure 33S). In all the cases of double tile systems where there is one type of constraint (*i.e.*, DTL(O,X) and DTL(X,O)), the sizes of the lattices were found to be of size **II**. A particularly interesting result is the structure formation of DTL(X,X). In the process of annealing, undesired tiles may bind to already formed DNA structures due to a phenomenon known as stacking. Due to the severe constraints enforced by the design of DTL(X,X), stacking seems to be completely suppressed which allows for a 1D wire of single DX tile width to form (Figure 3D).

**CONCLUSION**

In this study we have presented a scheme in which DNA lattices consisting of single and double DX tiles were constructed under hitherto unfeasible conditions. Conditions such as non-complementarity and geometrical incompatibility, which up to now had been construed as obstacles in the fabrication of DNA nanostructures, have been shown to foster formations of 2D lattices (STL(O,X), DTL(O,X), and DTL(X,O)) and 1D wires (DTL(X,X)). Furthermore, control of the coverage made possible by using these types of lattices may prove fruitful for DNA nanotechnological applications such as solar cells[20] and optical devices.[21] Although the exact mechanisms of non-complementary and geometrical incompatible bindings remain an open issue, future works of changing the number of complementary sticky end base pairs and changing the degree of helicity may shed more light on these matters. Nevertheless, these results show that even in constraint laden conditions, fabrication of DNA nanostructures is possible and much more robust than previously thought.

**MATERIALS AND METHODS**

Synthetic oligonucleotides were purchased from Integrated DNA Technologies (IDT, Coralville, IA) and purified by high performance liquid chromatography (HPLC). The details can be found on www.idtdna.com. Complexes were formed by mixing a stoichiometric quantity of each strand in a physiological buffer, 1×TAE/Mg$^{2+}$ [Tris-Acetate-EDTA (40 mM Tris, 1 mM EDTA (pH 8.0)) with 12.5 mM magnesium acetate]. The final concentrations of DNA were 100 nM, 200 nM, 400 nM, 600 nM, and 800 nM for single tile lattice samples and 200 nM for double tile lattice samples. For high temperature annealing, equimolar mixtures of strands were cooled slowly from 95 °C to 25°C by placing the AXYGEN-tubes in 3.5 L of boiled water in a styrofoam box for at least 48 hours to facilitate hybridization. To obtain the AFM images, 5 μL of samples were placed on freshly cleaved mica for 30 seconds and after which 45 μL 1×TAE/Mg$^{2+}$ buffer was pipetted onto the mica surface and another 5μL of 1×TAE/Mg$^{2+}$ buffer was dropped onto the AFM tip (Veeco Inc.). AFM images were taken by Nanoscope (Veeco Inc.) by liquid tapping.

*Supporting Information Available*: DNA base sequences of all DX tiles used in this experiment. Additional AFM images of STLs and DTLs. This material is available free of charge *via* the Internet at http://pubs.acs.org.

*Acknowledgements*: S. H. Park was supported by the SEOK CHUN Research Fund (2008/2009) at Sungkyunkwan University, Korea and the Basic Science Research Program through the National Research Foundation of Korea (NRF) funded by the Ministry of Education, Science and Technology (2010-0013294). Y. H. Kwon was supported by the Research Fund of HYU (HYU-2010-T) and Basic Science Research Program through the National Research Foundation of Korea (NRF) funded by the Ministry of Education, Science and Technology (KRF2010-0025620).

**FIGURE CAPTIONS**

**Figure 1.** Schematic diagrams of the types of double-crossover tiles used in the experiments. The body of the tile is represented as a square block and the sticky ends are represented as colored shapes. Same colored sticky ends represent bindings with Watson-Crick complementarity and complementary shapes represent geometrical compatibility. (A-D) Single tile lattices (STLs), with complementary and geometrically compatible sticky ends (STL(O,O), (A)), with complementarity and geometrical incompatibility (STL(O,X), (B)), with non-complementarity and geometrical compatibility (STL(X,O), (C)), and with non-complementarity and geometrical incompatibility (STL(X,X), (D)). Tiles with inverted labels represent STL tiles which are rotated along the longitudinal direction of the DX tile. (E-H) Double tile lattices (DTLs) with complementary and geometrically compatible sticky ends (DTL(O,O), (E)), with complementarity and geometrical incompatibility (DTL(O,X), (F)), with non-complementarity and geometrical compatibility (DTL(X,O), (G)), and with non-complementarity and geometrical incompatibility (DTL(X,X), (H)). For the DTLs in figures E-F, the yellow and red triangular sticky ends are bindings which satisfy complementarity and geometrical compatibility, also referred to as "common sets" in the text. The circles inside the bodies in figures F-H indicate hairpin structures.

**Figure 2.** Analysis of single tile lattices (STLs). (A-H) Atomic force microscope (AFM) images of STLs. AFM images of STLs under various constraints for DNA concentrations of (A-D) 400 nM and (E-H) 800 nM. The scan size of all images is 1 μm × 1 μm. (I) The concentration dependence of the average lattice sizes of STL(O,O) and STL(O,X). The dotted line indicates remnant patches of lattices. (J) The concentration dependence of the mica coverage of STL(O,O) and STL(O,X).

**Figure 3.** Analysis of double tile lattices (DTLs) (A-D) Atomic force microscope (AFM) images of DTLs. (A) AFM image of DTL(O,O) at 1 μm × 1 μm, (B) DTL(O,X) at 400 nm × 400 nm, (C) DTL(X,O) at 400 nm × 400 nm, and (D) DTL(X,X) at 2 μm × 2 μm. All images were taken at a DNA

concentration of 200 nM. (E-G) Cartoon of the possible lattice patterns for DTLs with one type of constraint, *i.e.*, DTL(O,X) and DTL(X,O). Yellow rectangles with an inner circle represent tiles with hairpin structures while blue rectangles represent ones without hairpin structures. Green bindings depict common sets and red bindings depict bindings with either non-complementarity or geometrical incompatibility. (H) Average lattice sizes of DTL(O,O), DTL(O,X), and DTL(X,O).

# FIGURES

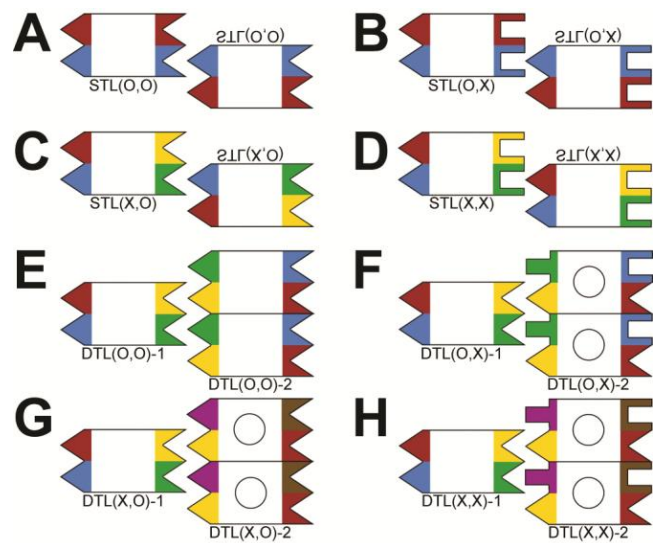

**Figure 1.**

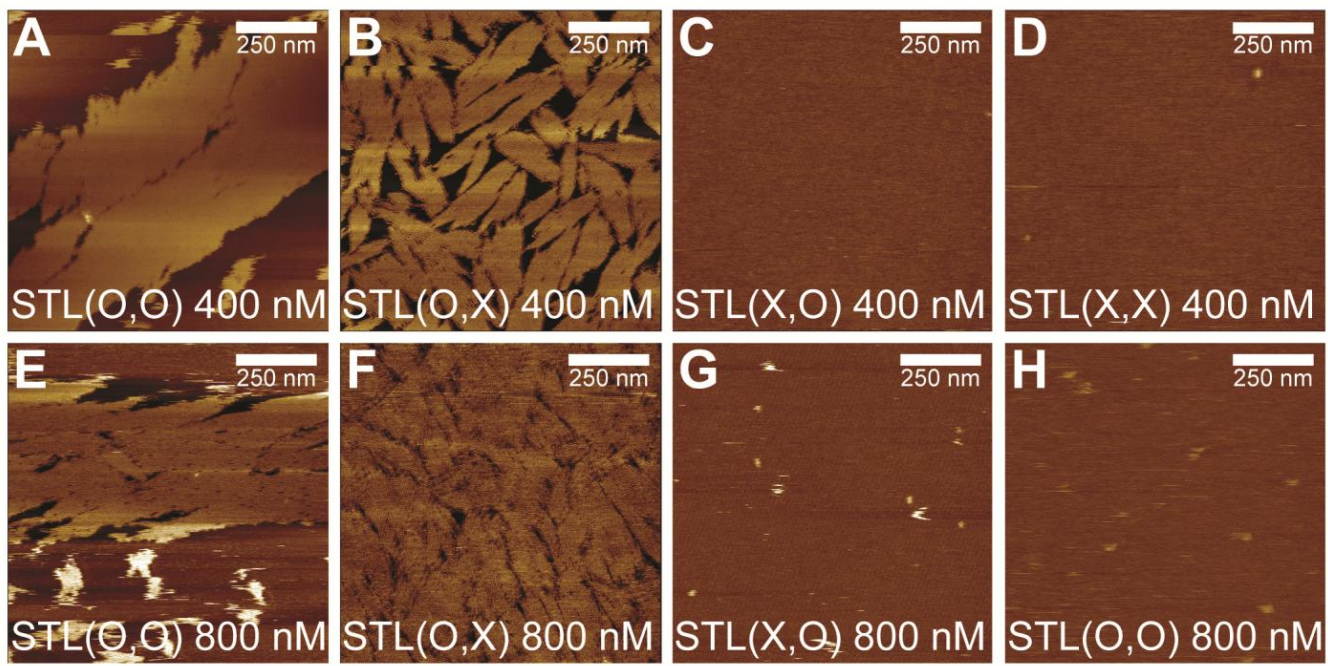

**Figure 2.**

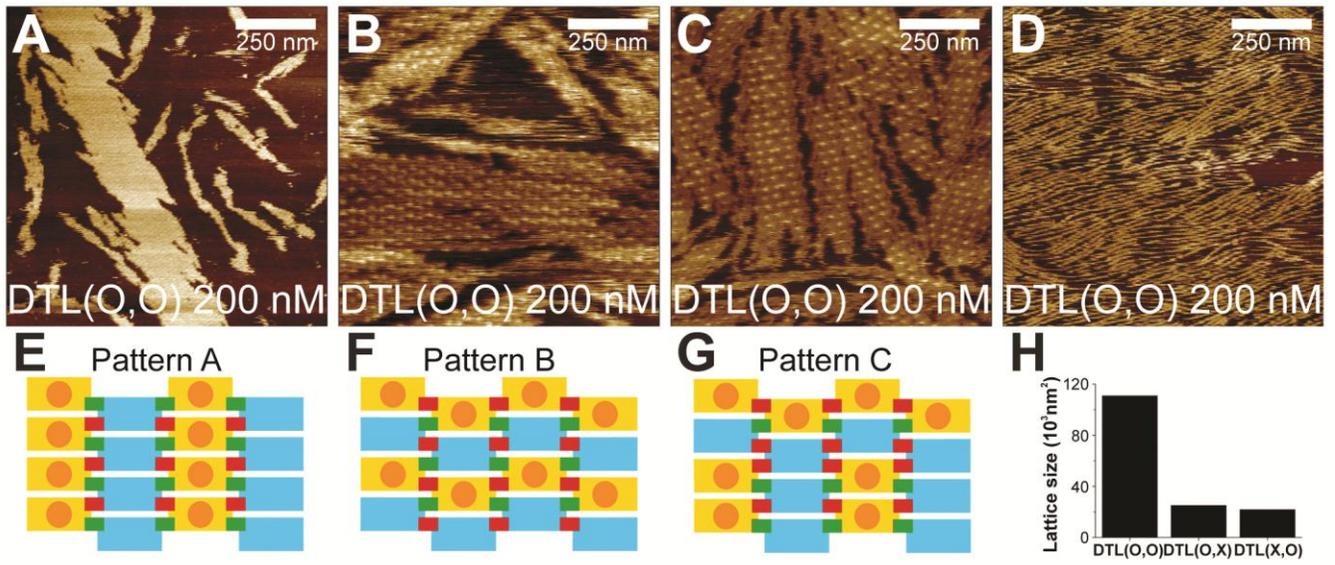

**Figure 3.**